\documentclass[a4paper,11pt]{article}
\usepackage{pos}

\newcommand{\mumu}{\mu^+\mu^-}
\newcommand{\pipi}{\pi^+\pi^-}

\newcommand{\mee}{e^+e^-}
\title{New BaBar studies of high-order radiation and the new landscape of data-driven HVP predictions of the muon $g-2$}
\author*[a]{Bogdan Malaescu}
\onbehalf{On behalf of the BaBar collaboration}
\affiliation[a]{ LPNHE, Sorbonne Université, Université Paris Cité, CNRS/IN2P3,\\
75252, Paris, France}

\emailAdd{malaescu@in2p3.fr}

\abstract{A measurement of additional radiation in $\mee \to \mumu \gamma$ and $\mee \to \pipi \gamma$ initial-state-radiation events is presented using the full $BaBar$ data sample. 
For the first time results are presented at next-to- and next-to-next-to-leading order, with one and two additional photons, respectively, for radiation from the initial and final states. 
The comparison with the predictions from \textsc{Phokhara} and \textsc{AfkQed} generators reveals discrepancies for the former in the one-photon rates and angular distributions. 
While this disagreement has a negligible effect on the $\mee \to \pipi (\gamma)$ cross section measured by $BaBar$, the impact on the KLOE and BESIII measurements is estimated and found to be indicative of significant systematic effects. 
The findings shed a new light on the longstanding deviation among the muon $g-2$ measurement, the Standard Model prediction using the data-driven dispersive approach for calculation of the hadronic vacuum polarization~(HVP), and the comparison with lattice QCD calculations.}

\FullConference{42nd International Conference on High Energy Physics (ICHEP2024)\\
18-24 July 2024\\
Prague, Czech Republic\\}

\begin{document}
\maketitle

\section{Introduction}

The dominant uncertainty of the theoretical prediction of the muon magnetic moment anomaly currently originates from the HVP contribution~($a_\mu^{had}$).
The determination of $a_\mu^{had}$ through the dispersive approach requires precise $\mee\to {\rm hadrons}$ data in the low mass region, the contribution of which is enhanced by the shape of the integration kernel.
In particular, the process $\mee\to\pipi(\gamma)$ provides $73\%$ of the HVP contribution and $70\%$ of its uncertainty squared.
Alternatively, one can use data from hadronic $\tau$ decays, with Isospin Breaking corrections.
The same data can be used to compute the HVP contribution for other quantities, like the restricted observable $a_\mu^{\rm win}$.

A large number of datasets were obtained through an energy scan approach,
the most precise ones being published by the CMD-2~\cite{CMD-2:2005mvb,Aulchenko:2006dxz,CMD-2:2006gxt}, SND~\cite{Achasov:2006vp,SND:2020nwa} and, more recently, CMD-3~\cite{CMD-3:2023alj} experiments.
The latter measurement represents one of the important novelties since the g-2 Theory Initiative White Paper~\cite{Aoyama:2020ynm}.
With the enhanced luminosities achieved at flavour factories, it became possible to employ the ISR method.
The KLOE collaboration employed a selection considering event topologies with a hard photon emitted by the initial-state leptons, plus two charged tracks reconstructed in the final state~\cite{KLOE:2008fmq,KLOE:2010qei,KLOE:2012anl,KLOE-2:2017fda}.
BaBar has pioneered the approach based on the ratio between the hadronic mass spectra and the $\mumu(\gamma)$ one, allowing to cancel many systematic uncertainties, yielding hence precise measured cross sections~\cite{BaBar:2009wpw,BaBar:2012bdw}.
BaBar provided the first "NLO" measurements, where a possible additional radiation is taken into account in the analysis, instead of being corrected a posteriori~(as done by other experiments).
In addition, the measured muon spectrum is compared with the NLO QED prediction, which represents an important cross check of the analysis.

These proceedings are based mainly on the studies for the HVP evaluation performed in Ref.~\cite{Davier:2019can}, the unique measurement of "(N)NLO" additional radiation~(up to two extra photons) performed by BaBar~\cite{BaBar:2023xiy} and the subsequent studies of the implications for other ISR measurements~\cite{Davier:2023fpl}.
We also discuss the comparison between the dispersive results for $a_\mu^{had}$ and the ones determined by the BMW lattice QCD collaboration~\cite{Borsanyi:2020mff}, the study of possible sources of differences~\cite{Davier:2023cyp} and the more recent merging of the two approaches~\cite{Boccaletti:2024guq}.

\section{Data combination and comparisons of mass spectra in the $\pipi$ channel}

In our studies we employ~(since 2009) a procedure for combining cross section data with arbitrary point spacing or binning, redistributed in a fine common binning using spline-based interpolations, as implemented in the HVPTools software~(see Refs.~\cite{Davier:2019can,Davier:2023fpl} and references therein).
For each narrow final bin a $\chi^2$ is minimized to get the average weights and test locally the level of agreement among the input measurements.
The average weights also account for the different bin sizes and point-spacing of measurements, in order to compare their precisions on the same footing.
This procedure has been validated through a closure test. 
It features full and realistic~(i.e.\ not too optimistic) treatment of uncertainties and correlations, between the measurements~(data points or bins) of a given experiment, between experiments and between different channels.
It also fully accounts for systematic tensions between experiments.

\begin{figure}[tb]
    \centering
    \includegraphics[width=7.0cm]{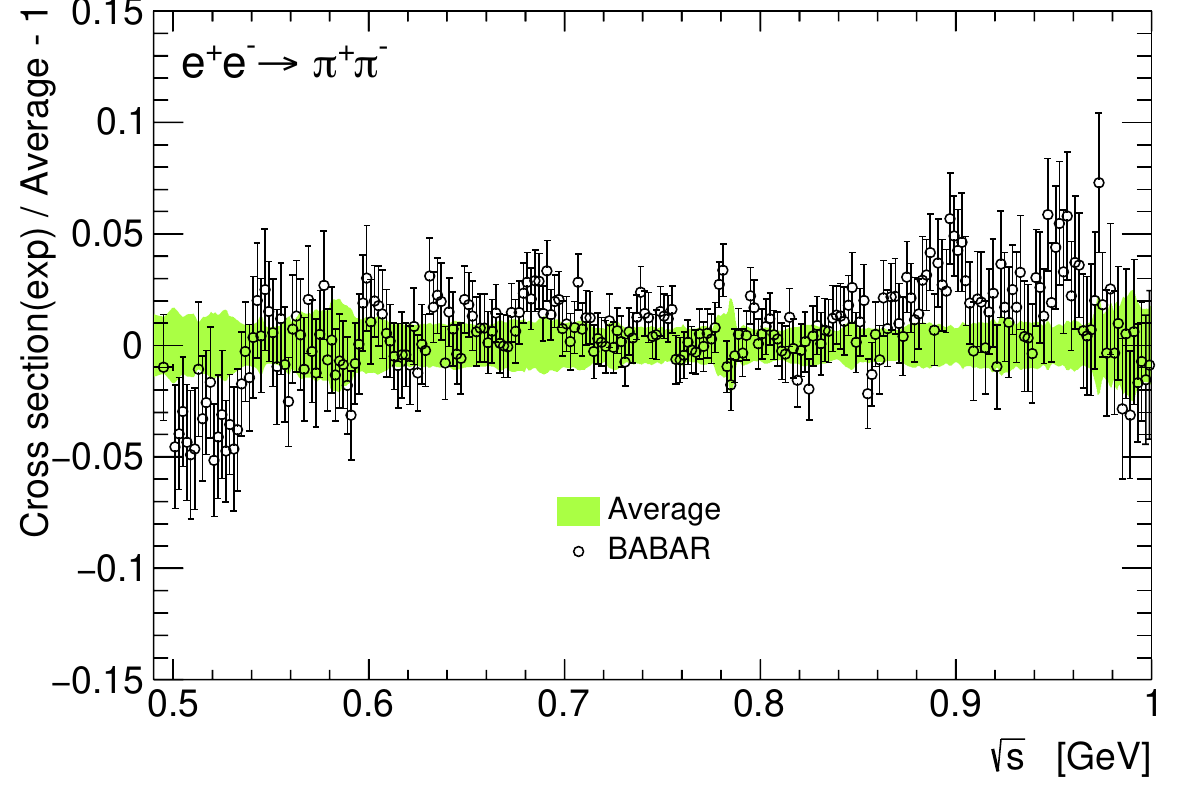}
    \hspace{0.3 cm}
    \includegraphics[width=7.0cm]{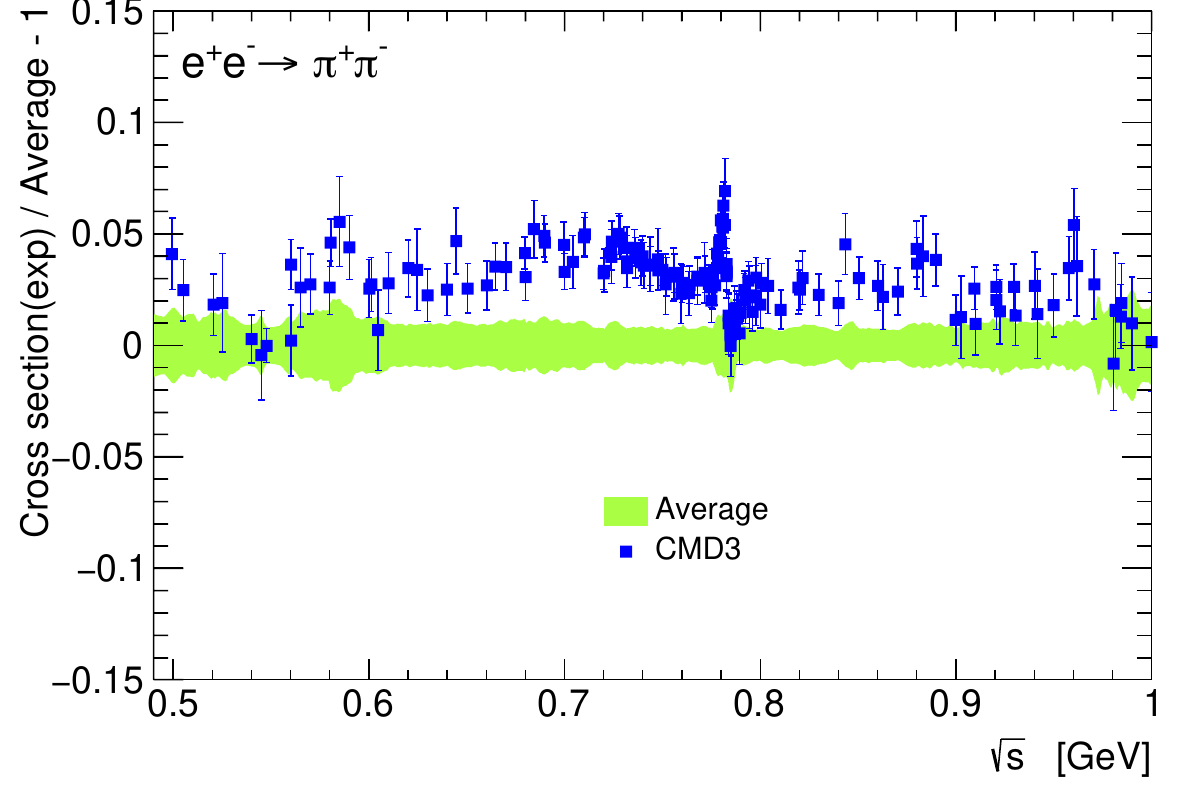} \\
    \includegraphics[width=7.0cm]{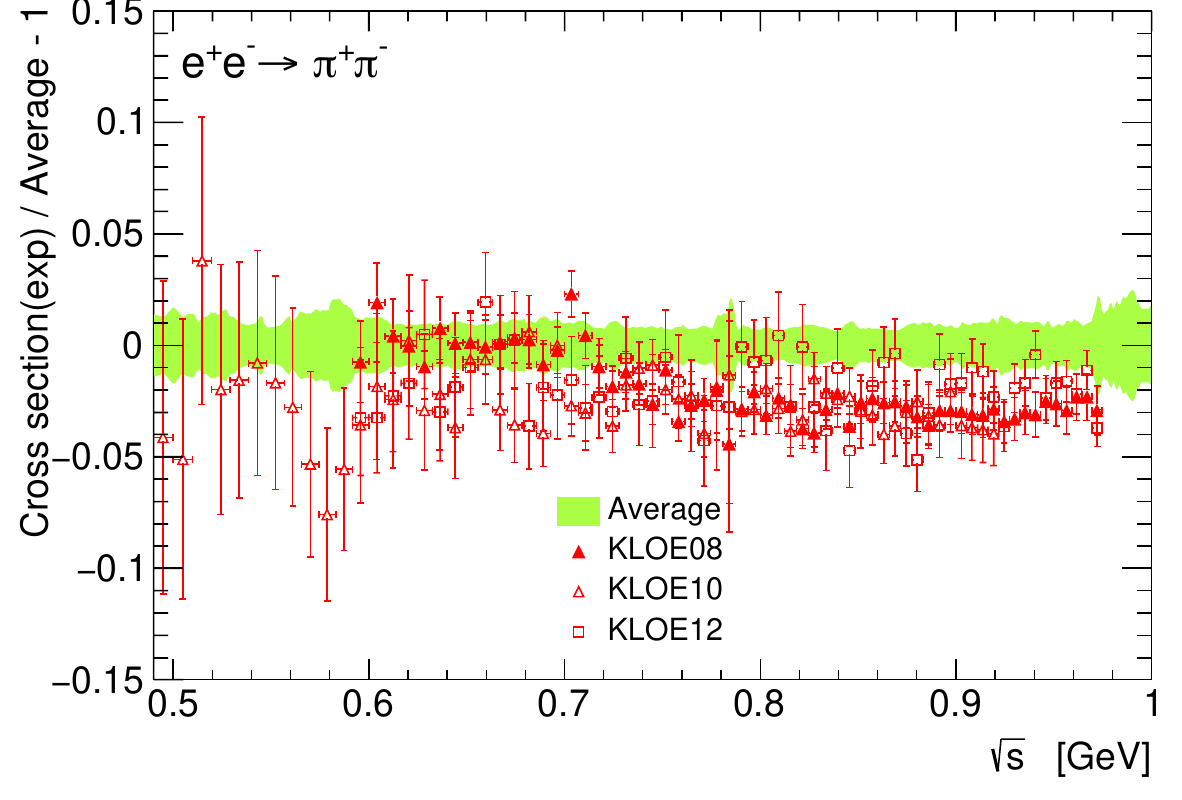}
    \hspace{0.3 cm}
    \includegraphics[width=6.5cm]{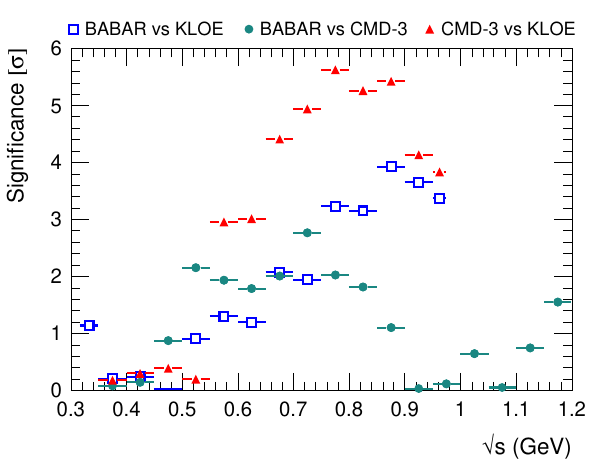}
  \caption{\small 
Comparison between the $\mee \to \pipi$ cross-section measurements from BaBar~\cite{BaBar:2009wpw,BaBar:2012bdw}~(top-left), CMD-3~\cite{CMD-3:2023alj}~(top-right), KLOE\,08~\cite{KLOE:2008fmq}, KLOE\,10~\cite{KLOE:2010qei}, KLOE\,12~\cite{KLOE:2012anl}~(bottom-left), and the HVPTools combination. Bottom-right: significance of the difference between pairs of the three most precise $\mee \to \pipi$ measurements for narrow energy intervals of $50~{\rm MeV}$ or less. Plots from Ref.~\cite{Davier:2023fpl}.}
\label{Fig:dataComparisons}
\end{figure}

The average is dominated by the most precise experiments (BaBar~\cite{BaBar:2009wpw,BaBar:2012bdw}, CMD3~\cite{CMD-3:2023alj}, KLOE~\cite{KLOE:2008fmq,KLOE:2010qei,KLOE:2012anl}, SND20~\cite{SND:2020nwa}), BaBar covering the full energy range of interest.
Taking the ratio between various measurements and this average, it is found that in the $[0.5;1]{\rm GeV}$ range the BaBar and SND20 data overlap rather well with the average, KLOE is systematically below it, while CMD3 is above~(see Fig.~\ref{Fig:dataComparisons}).
These tensions, especially between KLOE and CMD3, which provide the smallest and respectively largest cross-sections in the $\rho$ region, are also reflected by the enhanced values of $\chi^2/{\rm ndof}$.
For KLOE, slopes between the three various measurements are also observed.
These tensions were quantified through fits and found to be at the $2.5-3\sigma$ level~(see Ref.~\cite{Aoyama:2020ynm} and references therein).
Good agreement is found between BaBar and CMD3 at low and high mass~\cite{Davier:2023fpl}.

In order to further quantify these tensions, we compare the integrals computed in various restricted energy ranges, for individual experiments.
We determine the significance of the difference between pairs of experiments, taking into account the correlations of the uncertainties~(see Fig.~\ref{Fig:dataComparisons}, bottom-right).
The largest observed tensions are between CMD3 and KLOE, going beyond $5\sigma$ on the $\rho$ peak~\cite{Davier:2023fpl}.

The presence of these tensions among experimental measurements represents a clear indication of underestimated uncertainties.
This calls for a conservative uncertainty treatment in combination fits and in the determination of the averaging weights, as implemented in the DHMZ approach~\cite{Davier:2019can,Aoyama:2020ynm}.
These systematic tensions go well beyond the effects accounted through the local $\chi^2/{\rm ndof}$ rescaling.
This had already motivated the inclusion of the dominant BaBar-KLOE systematic by DHMZ, since the studies reported in Ref.~\cite{Davier:2019can}. 
However, the tensions are larger now and therefore require to understand their actual source.

\section{Impact of higher order photon emissions: a unique "(N)NLO" BaBar study}

The higher order photon emissions~(i.e.\ in addition to the hard ISR photon) are studied in-situ with BaBar data~\cite{BaBar:2023xiy}, using kinematic fits, in order to test the most frequently used Monte Carlo~(MC) generators.
Those are \textsc{Phokhara}, with full NLO matrix element for ISR and FSR, and \textsc{AfkQed}, including NLO and NNLO contributions, employing the collinear approximation for additional ISR.
The "(N)NLO" order counting in data and simulations is performed based on the number of additional photons in the final state, having the energy above some given threshold.

It is found that the rate of "NLO" small-angle ISR in \textsc{Phokhara} is higher than in data, while the data/MC ratios for large-angle photon emissions are consistent with unity~\cite{BaBar:2023xiy}.
An independent confirmation of the \textsc{Phokhara} problem has been provided by the measurement of the $\pi^+\pi^-\pi^0$ channel performed by the Belle-II Collaboration~\cite{Belle-II:2024msd}.
The "NNLO" contributions are also clearly observed in data, while they are missing in \textsc{Phokhara}.
At the same time, \textsc{AfkQed} provides a reasonable description of the rate and energy distributions for "(N)NLO" data.

It is important to note that the BaBar measurements~\cite{BaBar:2009wpw,BaBar:2012bdw} employ a loose selection, incorporating "NLO" and higher order radiation, minimising hence the dependence on MC simulations.
Other ISR measurements~\cite{KLOE:2008fmq,KLOE:2010qei,KLOE:2012anl} select "LO" event topologies and rely on \textsc{Phokhara} for hard "NLO" effects~(but lacking "NNLO"). 
These aspects have been further studied with fast simulations, questioning the KLOE systematic uncertainties~\cite{Davier:2023fpl}.

\section{A new perspective on $a_\mu^{had}$}

\begin{figure}[tb]
    \centering
    \includegraphics[width=7.0cm] {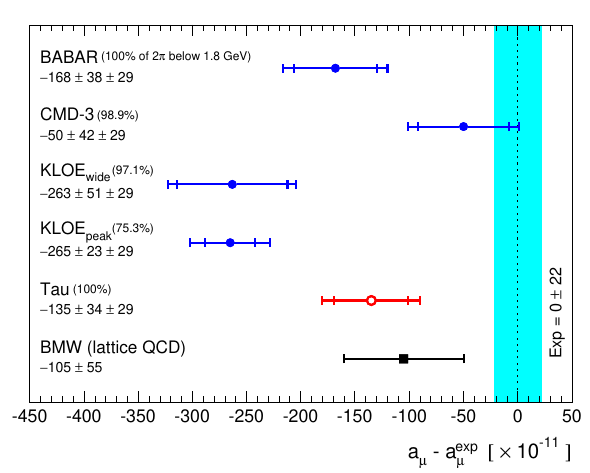}
    \hspace{0.3 cm}
    \includegraphics[width=7.0cm]{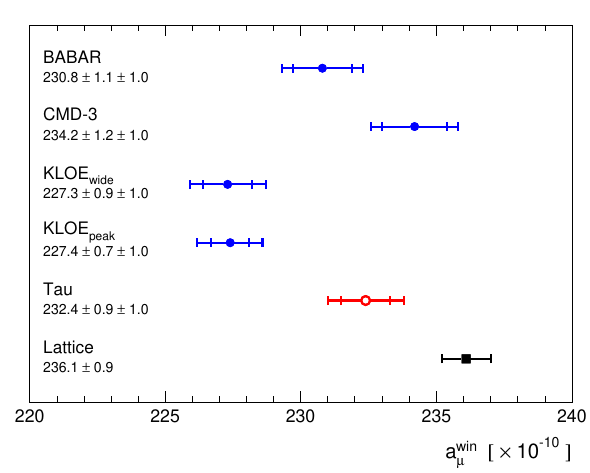}
  \caption{\small 
Dispersive predictions of $a_\mu$~(Left) and $a_\mu^{\rm win}$~(Right), based on various inputs in the $\pi\pi$ channel, compared with the BMW lattice QCD results and with the experimental measurement of $a_\mu$~(see text). Plots from Ref.~\cite{Davier:2023fpl}.}
\label{Fig:amu_amuWin}
\end{figure}

We consider $a_\mu$ and $a_\mu^{\rm win}$ calculations employing the dispersive approach, based on the most precise measurements available in the $\pipi(\gamma)$ channel, from BaBar~\cite{BaBar:2009wpw,BaBar:2012bdw}, CMD-3~\cite{CMD-3:2023alj} and KLOE~\cite{KLOE-2:2017fda}, as well as from hadronic $\tau$ decays~\cite{Davier:2013sfa}~(see Fig.~\ref{Fig:amu_amuWin}).
For KLOE, we consider both the full available range~(${\rm KLOE}_{\rm wide}$) and a restricted range of 0.6--0.975~${\rm GeV}$~(${\rm KLOE}_{\rm peak}$), where the data are most precise and KLOE's weight in the combination is largest.
These various $a_\mu^{had}$ integrals are completed with the combination of all the available measurements in the $\pi\pi$ channel, in order to cover the full mass range of interest, as well as with contributions from other hadronic channels, in view of the comparisons with the BMW lattice QCD result~\cite{Borsanyi:2020mff} and with the experimental measurement~\cite{Muong-2:2023cdq}.

The $\tau$-based HVP contribution is close to the values provided by BaBar and CMD-3~(see Fig.~\ref{Fig:amu_amuWin}).
Their combination~(which is $3.8~\sigma$ above ${\rm KLOE}_{\rm peak}$) is compatible with BMW for $a_\mu$, but a $2.9~\sigma$ tension persists for $a_\mu^{\rm win}$.
The BMW-based prediction is $1.8~\sigma$ below the experimental value of $a_\mu$.
Combining BaBar, CMD-3, $\tau$ (and BMW), a difference of $2.5~\sigma$ ($2.8~\sigma$) is found w.r.t. the experiment.
When including KLOE in the dispersive calculation, the difference w.r.t. the experiment becomes larger than $5~\sigma$.
Following these findings, tests of MC generators in the context of the KLOE analyses, with some focus on the rate and angular distributions of additional photons, and in-situ studies of the impact on the analysis are very much desirable~\cite{Davier:2023fpl}.

Studies based on the dispersive results preceding the CMD-3 measurement have shown that the difference w.r.t. the BMW result could be explained by an electromagnetic-current two-point function~(from lattice QCD) that is enhanced in the $[0.4; 1.5]~{\rm fm}$ range, or by enhancing measured hadronic spectra around (or in any larger interval including) the $\rho$-peak.~\footnote{ See also the contribution by Z. Fodor at this conference, on a new result with lattice QCD~(BMW) improvements and data-driven inputs~(DMZ) at large-t~\cite{Boccaletti:2024guq}. }
However, the required rescalings would be significantly larger than the quoted uncertainties.
The outcomes of these studies are stable within statistical and systematic uncertainties on lattice covariance matrices, studied here for the first time~\cite{Davier:2023cyp}.

\section{Conclusions}

We have an interesting, long standing, multifaceted problem and very important elements to solve the puzzle started to become available.
Future $\mee$ measurements will play a crucial role in this context.
In particular, an independent $\pi\pi$ measurement from BaBar, which does not employ particle identification in the determination of the cross-section~\cite{Michel17}, is expected soon.

As guiding ideas in these explorations, we need rigorous and realistic treatment of uncertainties and correlations at all levels, since underestimated uncertainties do not bring scientific progress and can put studies on wrong paths.
One also needs some caution about the interpretation of the significance:
while the $a_\mu$ measurement is statistics-dominated, the uncertainty of the prediction is limited by non-Gaussian systematic effects.
The studies for understanding differences between data-driven and lattice QCD approaches need to follow similar standards as the g-2 experiment, employing double-blinding, as we also do in the BMW-DMZ projects.

%\acknowledgements
\vspace{0.3cm}
{ \it The author would like to thank the organizers for the invitation to attend the conference and acknowledges the fruitful collaboration with Michel DAVIER, Andreas HOECKER, Anne-Marie LUTZ, Zhiqing ZHANG~(DHLMZ), and with the BMW lattice QCD collaboration. }

\end{document}